# Cation intermixing and ordering phenomenon in $M$–O layer of $M\text{Sr}_2\text{YCu}_2\text{O}_z$ ($M$ –1212) compounds with $M$ = Fe, Co, Al, and Ga: A neutron powder diffraction study

V.P.S. Awana, and E. Takayama-Muromachi

Superconducting Materials Center, National Institute for Materials Science, 1-1 Namiki, Tsukuba, Ibaraki 305-0044, Japan

S.K. Malik

Tata Institute of Fundamental Research, Homi Bhabha Road, Mumbai 400005, India

W.B. Yelon

Graduate Center for Materials Research, University of Missouri-Rolla, MO 65409, USA

M. Karppinen and H. Yamauchi

Materials and Structures Laboratory, Tokyo Institute of Technology, Yokohama 226-8503, Japan

Various $M\text{Sr}_2\text{YCu}_2\text{O}_z$ ($M$-1212 with $M$ = Fe, Co, Al and Ga) compounds have been synthesized through a solid-state reaction route. All these compounds are derived from the basic $\text{CuBa}_2\text{YCu}_2\text{O}_{7-\delta}$ structure. Rietveld structural refinement of the room-temperature neutron diffraction data for Fe-1212 reveals that nearly half of Fe remains at the $M$ site, while the other half goes to the Cu site in the $\text{CuO}_2$ plane. For the $M$ = Co, Al and Ga compounds, the basic unit cell gets doubled due to two distinct $M$ sites in the $M$-O layer and the space group $Ima2$, without any cation intermixing is found more appropriate. For the $M$ = Fe sample, magnetic ordering of Fe spins is seen from magnetization data below 70 K. In the case of the $M$ = Co compound, a shallow down-turn in magnetic susceptibility is seen below 150 K, indicating low-dimensional magnetic characteristics of the Co-O layer. The compounds with $M$ = Al and Ga show paramagnetic behaviour in their magnetization data down to 5 K. The oxygen content, established for each compound based on oxygen occupancies, refined from the neutron diffraction data comes close to 7.0 for each.

Key words: Neutron powder diffraction, $M$-1212, Oxygen ordering and Magnetic susceptibility.

## I. INTRODUCTION

High temperature superconductivity (HTSC) is believed to reside in the $CuO_2$ planes of all known HTSC compounds (see for instance [1], and references therein). The $CuO_2$ stacks are the integral part of all presently known HTSC superconductors. In $CuBa_2YCu_2O_{7-\delta}$ (Cu-1212), there are two different Cu sites, namely Cu(2) and Cu(1). Cu(2) resides in superconducting $CuO_2$ planes and Cu(1) in $CuO_{1-\delta}$ chains. Any alteration in $CuO_2$ superconducting stacks, even at the microscopic level, affects superconductivity drastically [1,2]. The $CuO_{1-\delta}$ chain acts as a redox layer and provides the mobile carriers to superconducting $CuO_2$ planes. Few successful attempts have been made in the past for complete replacement of $CuO_{1-\delta}$ by $M$-O layers with $M$ = Nb, Fe, Co, Ga, Al, etc. [3-6]. The problem of realising the perfect Cu-1212 structure without distributing the $M$ ions at both Cu(1) and Cu(2) sites along with the capability of $M$-O layers to dope mobile carriers in $CuO_2$ planes for achieving superconductivity had mainly been the focus of various studies in the past [3-8]. In the present contribution, we study $M$-1212 compounds, with $M$ = Fe, Co, Al and Ga to address the problem of $M$-cation intermixing with Cu and their structural peculiarities when compared with the parent Cu-1212 compound. For the synthesized $M$-1212 samples, magnetization data along with neutron powder diffraction (NPD) structural refinement details are presented and discussed.

## II. EXPERIMENT

The $M$-1212 samples with the composition $M$Sr$_2$YCu$_2$O$_z$ were synthesised through a solid-state reaction route starting from $Fe_2O_3$, $Co_2O_3$, $Al_2O_3$, $Ga_2O_3$, $SrO_2$, $Y_2O_3$, and CuO. Calcinations were carried out on mixed powders at 975 $^0$C, 1000 $^0$C, and 1010 $^0$C each for 24 hours with intermediate grindings. The pressed bar shape pellets were annealed in flowing oxygen at 1010 $^0$C for 40 hours and subsequently cooled slowly, over a span of another 20 hours, to room temperature. Magnetization measurements were carried out using a SQUID magnetometer (Quantum Design: MPMS-XL). Neutron diffraction patterns were obtained at room temperature at the University of Missouri Research Reactor Facility.

## III. RESULTS AND DISCUSSION

The $M$-1212 structure can be viewed as BaO(SrO)/CuO$_2$/RE/CuO$_2$/BaO(SrO) slabs being interconnected by a sheet of $M$ and O atoms with variable composition $M$O$_{1\pm\delta}$. The oxygen sites in the CuO$_2$ planes are identified as O(2) and O(3). Oxygen site in the Sr-O plane is called O(4). The RE plane is devoid of any oxygen. The oxygen sites in $M$O$_{1\pm\delta}$ layers are named O(1) (along b-axis) and O(5) (along a-axis). In tetragonal Cu-1212 structure O(2),O(3) and O(1),O(5) are indistinguishable as lattice parameters $a=b$. The neutron diffraction data on the present $M$-1212 samples have been analyzed by the Rietveld refinement procedure using the generalized structural analysis system (GSAS) program and the pseudo-Voigt peak shape functions. Neutron scattering cross-sections used are (in units of fm) 7.75 for Y, 7.202 for Sr, 5.81 for Cu, 0.3499 for Al, 0.249 for Co, 0.7288 for Ga and 9.54 for Fe.

The $M$-1212 compounds are isomorphic to CuBa$_2$YCu$_2$O$_{7-\delta}$ and hence the fitting of their NPD data was first tried in *P4/mmm* and *P/mmm* space groups, which though failed for $M$ = Co, Al and Ga. Not only were seen the difference in fitted and observed data intensities, but some small intensity NPD peaks could not be fitted at all, particularly the shoulder peak close to most intense peak near 2theta = 48$^0$. The fitting of the NPD data is later tried with the space group *Ima2*, considering two possible distinct $M$ sites in $M$-O layers. Figures 1 (a) and (b), depict the observed and fitted room temperature neutron diffraction pattern for Ga and Al based $M$-1212 compounds. Earlier careful SAED (selected area electron diffraction) and HRTEM (high-resolution transmission electron microscopy) studies on Al-1212 compound revealed a more complex structure [9]. A superstructure due to the existence of two types of symmetry-related zigzag Al-O$_4$ chains along the a-direction was found resulting in $a = 2a_t$, b = $4a_t$ and c = $2c_t$, where $a_t$ and $c_t$ are cell parameters for *P4/mmm* space group. This may be true in case of other $M$-O$_4$ chains also, hence careful micro-structural studies on various $M$-1212 compounds are warranted. Our NPD data in Figs. 1 (a) and (b) indicate that for $M$ = Al, Ga and Co (for Co-1212, the NPD pattern is given in ref.10) the space group of crystallization is *Ima2*. The general situation is similar to that observed in micro-structural studies, as the doubling of the primitive $M$-1212 cell being observed but not the superstructures due to zigzag $M$-O$_4$ chains in $a$ and $b$ directions. This may happen due to the average of different domains generated on going from primitive *M*-1212 cell to higher one in *Ima2* space group. Superstructures in *a*, *b* directions were seen in single crystal XRD data of Al-1212 [11], thus confirming the SAED and HRTEM findings on average scale also. In our NPD data on bulk polycrystalline samples, we could not observe the average *a*, *b* direction superstructures. Only doubling of unit cell and the change of space group to *Ima2* are concluded. Full substitution of Cu(1) in Cu-1212 by higher-valence *M* ions like Ga, Co, or Al invokes for two crystallographically distinct *M* sites due to change in oxygen environment resulting in changed space group. In space group *Ima2*, the lattice parameters obtained are: $a$ = 22.8048(23) Å, $b$ = 5.3954(6) Å, $c$ = 5.4834(6) Å for Ga-1212; $a$ = 22.1998(11) Å, $b$ = 5.4649(7) Å, $c$ = 5.46338(6) Å for Al-1212, and $a$ = 22.7987(18) Å, $b$ = 5.4515(5) Å, $c$ = 5.4089(5) Å for Co-1212. Here it is worth mentioning that to compare *P4/mmm* with *Ima2* the lattice parameter *a* gets interchanged with *c* and the factor of $\sqrt{2}$ is applied to both *a* and *b*. It is interesting to note that in case of Ga-1212 and Co-1212 clear orthorhombic distortions are seen, but not for Al-1212. In fact Al-1212 is more close to a primitive *P4/mmm* tetragonal cell with $a = b = 4.46/\sqrt{2}$ Å and $c$ = 22.199/2 Å. The structure of Al-1212 is discussed in detail in ref.10, and is more correctly to be treated in space group *P4/mmm*, with superstructures due to zigzag nature of $AlO_4$ chains in both *a, b* directions. Zigzag nature of $AlO_4$ chains separates them microscopically from $CoO_4$ and $GaO_4$. Here for direct comparison, we compare Al-1212 data with Ga/Co-1212 in space group *Ima2*. The structural and other parameters obtained from the fit for those *M*-1212 compounds, which crystallize in space group *Ima2* are listed in Table 1. In brief, two types of O(1) are observed in $MO_{1-\delta}$ chains with different coordinates resulting in O(1a) and O(1b), and thus doubling the unit cells. The overall oxygen content for the present *M*-1212 samples with *M* = Al, Ga, and Co, being determined from various oxygen occupancies is close to 7.0.

In the case of Fe-1212, we observed an altogether different situation; the Fe ions were found to distribute equally between Cu(1) and Cu(2) sites in the *M*-1212 structure, and the crystallization of it took place in *P4/mmm* structure, see ref.12. It was seen that nearly 47% of Fe occupies the Cu(2) site in $CuO_2$ planes and, as a result, same amount of Cu lies at the Cu(1) site, otherwise designated solely for Fe. Overall oxygen content, as determined from oxygen occupancies, was 7.30(2). The resultant formula for the compound is $(Fe_{0.53}Cu_{0.47})_{1}Sr_2Y(Cu_{0.53}Fe_{0.47})_{2}O_{7.30}$. The intermixing of Fe and Cu does not allow the long-range oxygen ordering in $(Fe,Cu)O_{1-\delta}$ layers and hence no two different Fe sites result in doubling of unit cells and the change in space group to *Ima2*.

Figures 2 (a), (b) and (c) show plots of magnetic susceptibility ($\chi$) vs. temperature (T) data for the M = Fe, Co and Al samples measured in an applied field of 100 Oe in temperature range of 5 K to 300 K in both field-cooled (FC) and zero-field-cooled (ZFC) states. Clear branching of FC and ZFC magnetizations is seen at around 78 K for the Fe-1212 sample with a further upturn in both ZFC and FC susceptibility at low T. The isothermal magnetization measurements at 5, 50 and 100 K, show no hysteresis loops at any of these temperatures (see inset of Fig. 2(a)). Temperature dependent NPD and Mössbauer studies had earlier been inconclusive on the nature of exact type of ordering of Fe spins in Fe-1212 [13]. For Co-1212, the low-dimensional like magnetic ordering of Co spins is seen from the behavior of the magnetic susceptibility below around 150 K, with a slight upturn at low temperatures, see Fig 2(b). In Fig. 2(c), shown is the magnetization data of Al-1212, which is basically paramagnetic down to 5 K. This may stem from the divalent Cu in under-doped $CuO_2$ planes, as no other cation is magnetic in this compound. The inverse of susceptibility *versus* temperature plot is shown in the inset of the figure, which is linear. Roughly estimated value of paramagnetic moment is 0.56 $\mu_B$ *per* formula unit. Fitting the data to Curie-Weiss paramagnetic equation for an under-doped non-superconducting HTSC compound might not be though feasible, because in an under-doped HTSC compound, the Cu spins might have ordered AFM above room temperature. HTSC cuprates phase diagrams are not yet clear at the point of quench of the AFM Cu state and the appearance of superconductivity. Furthermore, though the Cu may not be ordered AFM in the under-doped regime, it may carry some paramagnetic moment with low $T_c$ superconductivity. Our currently studied (Al,Ga)-1212 compounds are not superconducting and show paramagnetic behavior in their magnetic susceptibility down to 5 K. Though the NPD data is clean, the presence of very small paramagnetic impurity cannot be completely ruled out.

**ACKNOWLEDGEMENT**

V.P.S.A. acknowledges the support of Prof. E. Takayama-Muromachi for providing him with the NIMS postdoctoral fellowship to work in NIMS.

**FIGURE CAPTIONS**

Figure 1. Observed and fitted neutron diffraction patterns for (a) GaSr$_2$YCu$_2$O$_{7\pm\delta}$, and (b) AlSr$_2$YCu$_2$O$_{7\pm\delta}$.

Figure 2. Magnetic susceptibility ($\chi$) vs. temperature (T) plot for (a) FeSr$_2$YCu$_2$O$_{7\pm\delta}$, (b) CoSr$_2$YCu$_2$O$_{7\pm\delta}$, and (c) AlSr$_2$YCu$_2$O$_{7\pm\delta}$, measured both in FC (Field-cooled) and ZFC (Zero- field-cooled) states in an applied field of 100 Oe. Inset of (a) shows the M (magnetization) vs. H (applied magnetic field) behaviour for FeSr$_2$YCu$_2$O$_{7\pm\delta}$, and inset of (c) shows the inverse of magnetic susceptibility ($\chi^{-1}$) vs. T plot for AlSr$_2$YCu$_2$O$_{7\pm\delta}$.

**Table 1.** Refined structural parameters for $M$Sr$_2$YCu$_2$O$_7$ compounds ($M$ = Al, Ga and Co), including the atomic coordinates, occupancies, and thermal parameters (U$_{iso}$). Space group *Ima2*, lattice parameters are $a$ = 22.8048(23) Å, $b$ = 5.3954(6) Å, $c$ = 5.4834(6) Å for Ga-1212; $a$ = 22.1998(11) Å, $b$ = 5.4649(7) Å, $c$ = 5.46338(6) Å for Al-1212, and $a$ = 22.7987(18) Å, $b$ = 5.4515(5) Å, $c$ = 5.4089(5) Å for Co-1212. The error bars for all coordinates positions and variable occupancies are only after third decimal place.

| Atom | Ga-1212, *Ima2* R$_{wp}$= 5.27 % | Al-1212, *Ima2* R$_{wp}$= 4.40 % | Co-1212, *Ima2* R$_{wp}$= 4.25 % |
|---|---|---|---|
| Y (x, y, z) | 0.0, 0.0, 0.0 | 0.0, 0.0, 0.0 | 0.0, 0.0, 0.0 |
| Occupancy | 1.00 | 1.00 | 1.00 |
| 100 x U$_{iso}$(Å$^2$) | 1.4(1) | 1.5(1) | 1.17(10) |
| Sr (x, y, z) | 0.3492, 0.0104, 0.0063 | 0.3454, 0.0045, 0.0230 | 0.34818, 0.0045, 0.0041 |
| Occupancy | 2.0 | 2.0 | 2.0 |
| 100 x U$_{iso}$(Å$^2$) | 0.9(5) | 1.4(5) | 1.27(7) |
| M (x, y, z) | 0.2500, 0.5101, 0.0698 | 0.2500, 0.4636, 0.0047 | 0.2500, 0.5370, 0.0667 |
| Occupancy | 1.00 | 1.00 | 1.00 |
| 100 x U$_{iso}$(Å$^2$) | 3.3(8) | 2.2(8) | 3.1(6) |
| Cu (x, y, z) | 0.4270, 0.0119, 0.4939 | 0.4243, 0.0038, 0.5060 | 0.42683, 0.0053, 0.4947 |
| Occupancy | 2.0 | 2.0 | 2.0 |
| 100 x U$_{iso}$(Å$^2$) | 1.8 (5) | 1.3 (5) | 1.2 (6) |
| O(1a) (x, y, z) | 0.2500, 0.6510, 0.3517 | 0.2500, 0.7096, 0.2954 | 0.2500, 0.6489, 0.3547 |
| Occupancy | 0.3983 | 0.8615 | 0.3400 |
| 100 x U$_{iso}$(Å$^2$) | 1.50(10) | 1.40(10) | 1.15(29) |
| O(1b) (x, y, z) | 0.2500, 0.6091, 0.5858 | 0.2500, 0.4783, 0.5624 | 0.2500, 0.5906, 0.5932 |
| Occupancy | 0.5923 | 0.1925 | 0.6600 |
| 100 x U$_{iso}$(Å$^2$) | 1.50(14) | 1.40(14) | 1.50(14) |
| O(2) (x, y, z) | 0.4405, 0.7355, 0.2505 | 0.4315, 0.7421, 0.2564 | 0.43560, 0.7422, 0.2458 |
| Occupancy | 2.00 | 2.00 | 2.00 |
| 100 x U$_{iso}$(Å$^2$) | 1.20(18) | 2.00(18) | 1.86(19) |
| O(3) (x, y, z) | 0.4327, 02481, 0.7425, | 0.4314, 0.2440, 0.7677 | 0.43615, 0.2424, 0.7479 |
| Occupancy | 2.00 | 2.00 | 2.00 |
| 100 x U$_{iso}$(Å$^2$) | 1.80(10) | 2.10(10) | 1.30(20) |
| O(4)x, y, z | 0.3224, 0.4728, 0.0306 | 0.3193, 0.4935, 0.0939 | 0.32413, 0.4783, 0.0310 |
| Occupancy | 2.00 | 2.00 | 2.00 |
| 100 x U$_{iso}$(Å$^2$) | 1.90(20) | 4.50(20) | 1.37(12) |

*Fig.1 (a) Awana et al. 47th MMM; MS FU-14*

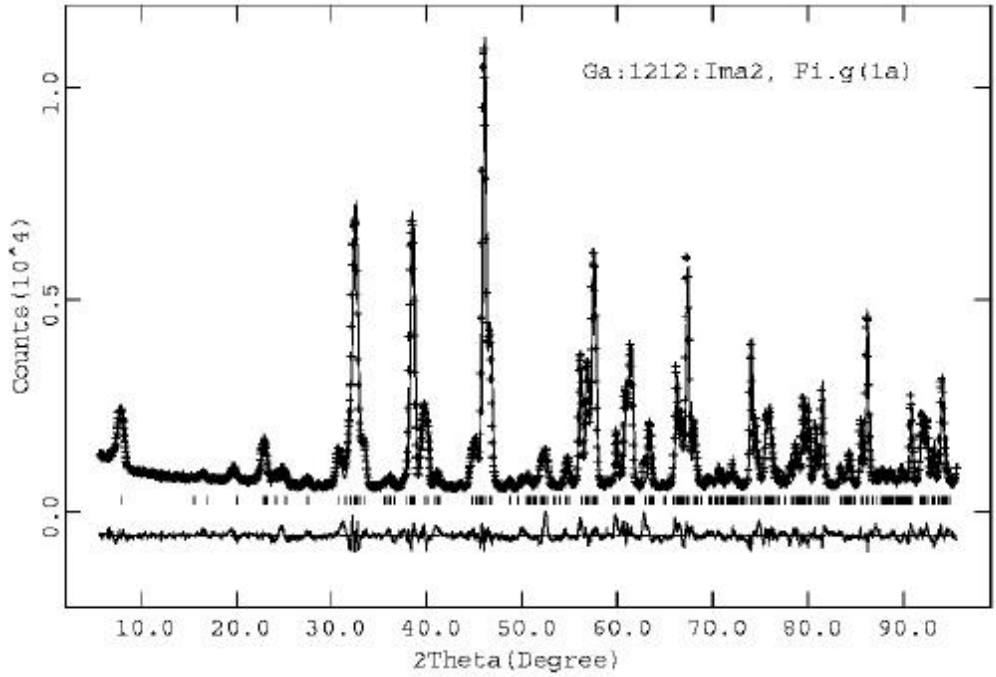

*Fig.1 (b) Awana etal. 47th MMM; MS FU-14*

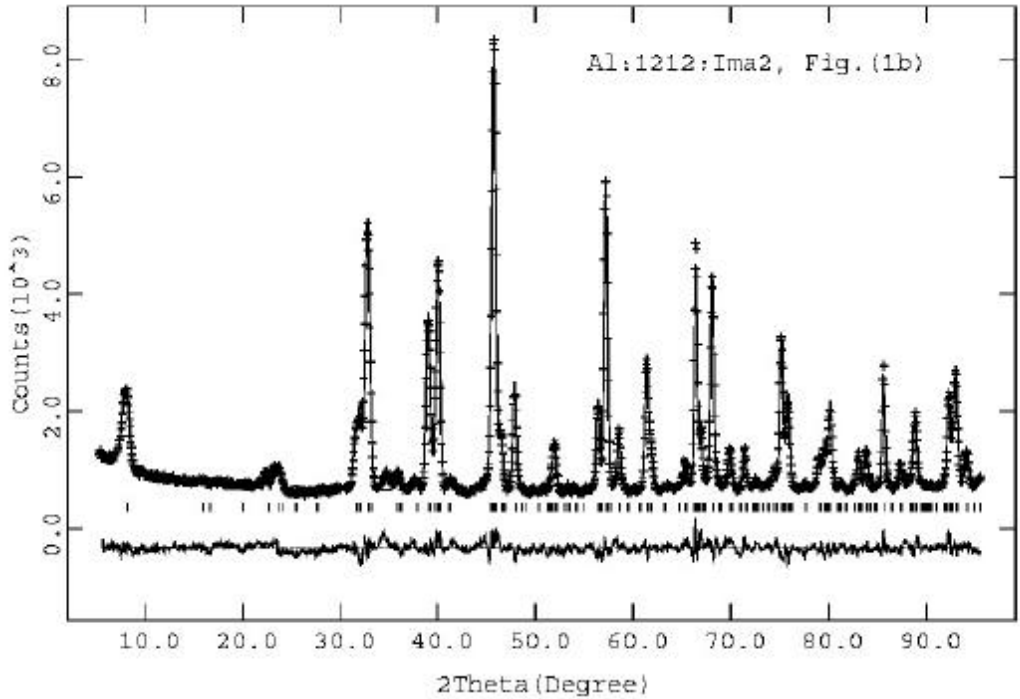

*Fig.2 (a) Awana et al. 47th MMM; MS FU-14*

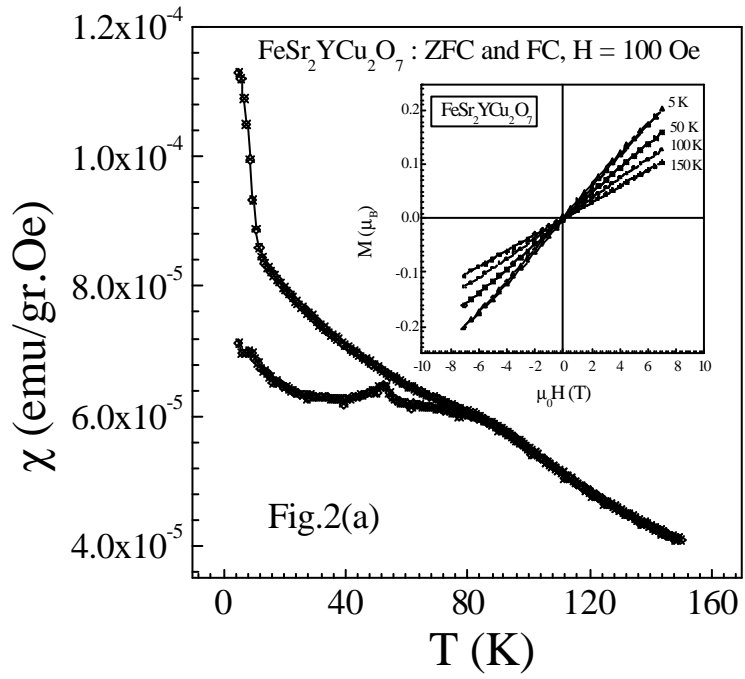

*Fig.2 (b) Awana et al. 47th MMM; MS FU-14*

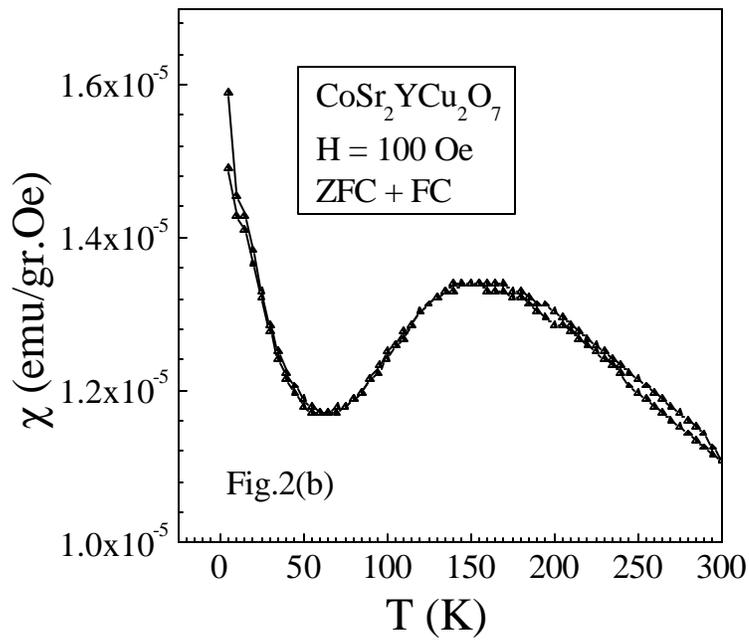

*Fig.2 (c) Awana et al. 47th MMM; MS FU-14*

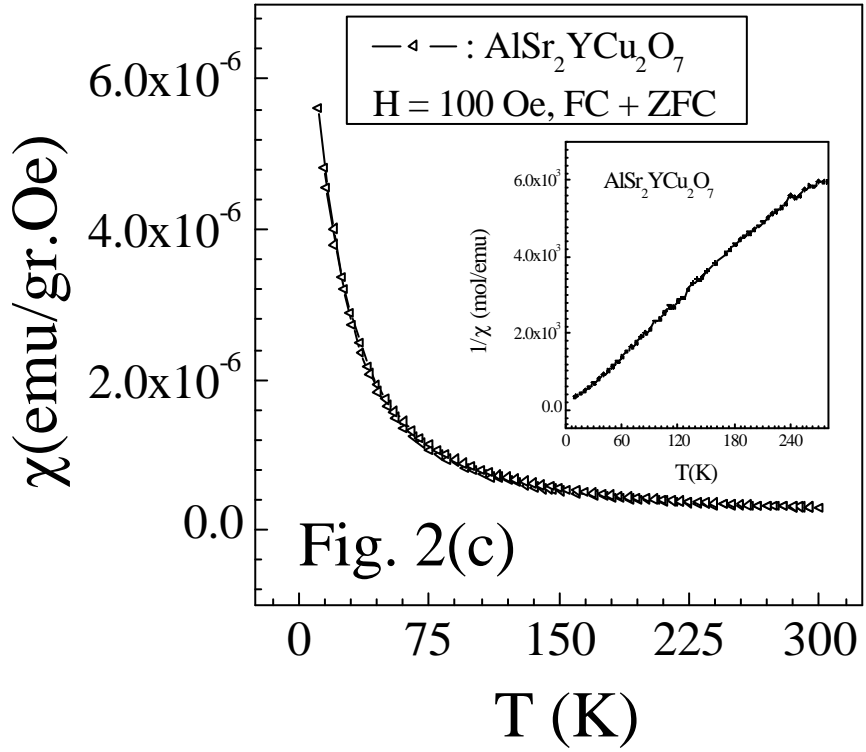